\documentclass[prl,aps,twocolumn]{revtex4}

\usepackage{psfrag}
\usepackage{graphicx}
\usepackage{color}
\usepackage{dcolumn}
\usepackage{latexsym,amsfonts}
\usepackage{bm}
\usepackage{amssymb}
\begin{document}

\title{On the influence of the magnetic field of the GSI experimental
storage ring on the time--modulation of the $EC$--decay rates of the
H--like mother ions} \author{M. Faber$^{a}$\thanks{E-mail:
faber@kph.tuwien.ac.at}, A. N. Ivanov${^{\,a,b}}$, P. Kienle$^{b,c}$,
M. Pitschmann${^a}$, N. I. Troitskaya$^{e}$}
\affiliation{${^a}$Atominstitut der \"Osterreichischen
Universit\"aten, Technische Universit\"at Wien, Wiedner Hauptstrasse
8-10, A-1040 Wien, \"Osterreich} \affiliation{${^b}$Stefan Meyer
Institut f\"ur subatomare Physik \"Osterreichische Akademie der
Wissenschaften, Boltzmanngasse 3, A-1090, Wien,
\"Osterreich}\affiliation{${^c}$Excellence Cluster Universe Technische
Universit\"at M\"unchen, D-85748 Garching, Germany} \affiliation{
$^e$State Polytechnic University of St. Petersburg, Polytechnicheskaya
29, 195251, Russia} \email{ivanov@kph.tuwien.ac.at}

\date{\today}

\begin{abstract}
  We investigate the influence of the magnetic field of the
  Experimental storage ring (ESR) at GSI on the periodic
  time--dependence of the orbital K--shell electron capture decay
  $(EC$) rates of the H--like heavy ions. We approximate the magnetic
  field of the ESR by a uniform magnetic field. Unlike the assertion
  by Lambiase {\it et al.}, arXiv: 0811.2302 [nucl--th], we show that
  a motion of the H--like heavy ion in a uniform magnetic field cannot
  be the origin of the periodic time--dependence of the $EC$--decay
  rates of the H--like heavy ions. 
\end{abstract}

\maketitle

\section{Introduction}

Recently Litvinov {\it et al.} \cite{GSI2} have observed that the
K--shell electron capture ($EC$) decay rates of H--like ${^{140}}{\rm
Pr}^{58+}$ and ${^{142}}{\rm Pm}^{60+}$ ions
\begin{eqnarray}\label{label1}
&&{^{140}}{\rm Pr}^{58+} \to {^{140}}{\rm Ce}^{58+} + \nu_e,\nonumber\\
&&{^{142}}{\rm Pm}^{60+} \to {^{142}}{\rm Nd}^{60+} + \nu_e
\end{eqnarray}
have an unexpected periodic time modulation of exponential decay
curves. The rates of the number $N^{EC}_d$ of daughter ions
${^{140}}{\rm Ce}^{58+}$ and ${^{142}}{\rm Nd}^{60+}$
\begin{eqnarray}\label{label2}
  \frac{dN^{EC}_d(t)}{dt} = \lambda_{EC}(t)\, N_m(t),
\end{eqnarray}
where $N_m(t)$ is the number of the H--like mother ions ${^{140}}{\rm
Pr}^{58+}$ or ${^{142}}{\rm Pm}^{60+}$ \cite{GSI2} and $\lambda^{(\rm
H)}_{EC}(t)$ is the $EC$--decay rate, are periodic functions, caused
by a periodic time--dependence of the $EC$--decay rates
\begin{eqnarray}\label{label3}
  \lambda_{EC}(t) = \lambda_{EC}\,(1 + a^{EC}_d\, \cos(\omega_{EC}t  +
\phi_{EC}))
\end{eqnarray}
with periods $T_{EC} = 2\pi/\omega_{EC} = 7.06(8)\,{\rm s}$ and
$T_{EC} = 2\pi/\omega_{EC} = 7.11(22)\,{\rm s}$ for the $EC$--decays
of ${^{140}}{\rm Pr}^{58+}$ or ${^{142}}{\rm Pm}^{60+}$, respectively,
amplitudes $a^{EC}_d \simeq 0.20$ and phases $\phi_{EC}$. Below such a
periodic time--dependence we call the ``GSI oscillations''.

Recently \cite{GSI4}--\cite{GSI6} the decay rate of the $EC$--decay
${^{122}}{\rm I}^{52+} \to {^{122}}{\rm Te}^{52+} + \nu_e$ with a
period of the time--modulation $T_{EC} = 6.11(3)\,{\rm s}$ has been
observed. As has been pointed out in \cite{GSI4}--\cite{GSI6}, the
periods of the time--modulation of the H--like heavy ions obey the
$A$-scaling: $T_{EC} = A/20\,{\rm s}$, where $A$ is a mass number of
the mother H--like heavy ions.

In the articles \cite{Ivanov5} (see also \cite{Ivanov2,Ivanov3}) we
have proposed an explanation of the periodic time--dependence of the
$EC$--decay rates as an interference of two neutrino mass--eigenstates
$\nu_1$ and $\nu_2$ with masses $m_1$ and $m_2$, respectively. The
period $T_{EC}$ of the time--dependence has been related to the
difference $\Delta m^2_{21} = m^2_2 - m^2_1$ of the squared neutrino
masses $m_2$ and $m_1$ as follows
\begin{eqnarray}\label{label4}
\omega_{EC} = \frac{2\pi}{T_{EC}} = \frac{\Delta m^2_{21}}{2\gamma
M_m},
\end{eqnarray}
where $\gamma M_m$ is the energy of the H--like mother ion with mass
$M_m$ in the ESR and $\gamma = 1.43$ is a Lorentz factor \cite{GSI2}.
In a subsequent analysis we also showed that the $\beta^+$--branches
of the decaying H--like heavy ions do not show time modulation,
because of the broad continuous energy spectrum of the neutrinos
\cite{Ivanov5}. This agrees well with the experimental data
\cite{GSI4}--\cite{GSI6}.

According to the atomic quantum beat experiments and theory
\cite{QB1,QB2}, the interpretation of the ``GSI oscillations'',
proposed in \cite{Ivanov5}--\cite{Ivanov3}, bears similarity with
quantum beats of atomic transitions, when an excited atomic eigenstate
decays into a coherent state of two (or several) lower lying atomic
eigenstates. In the case of the $EC$--decay one deals with a
transition from the initial state $|m\rangle$ to the final state
$|d\,\nu_e\rangle$, where the electron neutrino is a coherent
superposition of two neutrino mass--eigenstates with the energy
difference equal to $\omega_{21} = \Delta m^2_{21}/2 M_m$ related to
$\omega_{EC}$ as $\omega_{EC} = \omega_{21}/\gamma$.

As has been pointed out in \cite{GL08}, a motion of the H--like mother
ion in the magnetic field of the ESR can be the origin of the ``GSI
oscillations'' \cite{GSI2}. In this letter we investigate the
influence of the magnetic field of the ESR at GSI making a consistent
calculation of the $EC$--decay rate by using the weak interaction
Hamilton operator and taking into account a motion of the mother
H--like ion in the magnetic field. For simplicity we approximate the
magnetic field of the ESR at GSI by a constant magnetic field $\vec{B}
= B_0\,\vec{e}_z$ directed perpendicular the plane of the
ESR. However, we neglect also a possible quantisation of the energy of
the mother H--like in the constant magnetic field \cite{LL07} and take
into account only the interaction of a spin of the mother H--like ion
with a constant magnetic field.  We show that such a spin--rotation
coupling of the H--like heavy ions cannot be responsible for the
periodic time--dependence of the $EC$--decay rates, measured at GSI
\cite{GSI2}--\cite{GSI6}.

The Hamilton of the interaction of the H--like ions with a magnetic
field $\vec{B} = B_0\,\vec{e}_z$ we define as \cite{JJ83}
\begin{eqnarray}\label{label5}
\hspace{-0.3in}H_{\vec{B}} &=& 2\,\Big(a_e + \frac{1}{\gamma}\Big)\,
\mu_B\vec{s}\cdot \vec{B}\nonumber\\
 \hspace{-0.3in}&-& \Big(g_I - \frac{2 Z m_p}{M_I}\Big(1 -
\frac{1}{\gamma}\Big)\Big)\,\mu_N\,\vec{I}\cdot \vec{B},
\end{eqnarray}
where $\vec{s} = \frac{1}{2}\,\vec{\sigma}$ and $\vec{I}$ are
operators of spins of the electron and the mother nucleus with
eigenvalues $s = \frac{1}{2}$ and $I = 1$; $a_e = (g_e - 2)/2$ is the
anomalous magnetic moment of the bound electron with $g_e$ equal to
\cite{BS57}--\cite{HFS3}
\begin{eqnarray}\label{label6}
\hspace{-0.3in}\frac{1}{2}\,g_e &=& 1 + \frac{2}{3}\,(\sqrt{1 -
  (\alpha Z)^2} - 1)\nonumber\\ \hspace{-0.3in}&+&
\frac{\alpha}{\pi}\Big(\frac{1}{2} + \frac{1}{12}\,(\alpha Z)^2 +
\frac{7}{2} \,(\alpha Z)^4 + \ldots \Big),
\end{eqnarray}
where $Z = 59$ for the H--like heavy ion ${^{140}}{\rm Pr}^{58+}$;
$g_I = \mu_I/I$ and $M_I$ are the anomalous magnetic moment of the
nucleus with spin $I$ and the mass. For the nucleus ${^{140}}{\rm
Pr}^{59+}$ they are equal to $g_I = 2.5$ \cite{GSI1} and $M_I =
130324.46\,{\rm MeV}$. Then, $\mu_B = e/2m_e = 5.788\times
10^{-5}\,{\rm eV\,T^{-1}}$ and $\mu_N = e/2m_p = 3.152\times
10^{-8}\,{\rm eV\,T^{-1}}$ are the Bohr and nuclear magnetons
\cite{PDG08}; $\gamma = 1.43$ is a Lorentz factor of a motion of a
H--like heavy ion in the ESR at GSI \cite{GSI2}, the value of the
magnetic field is $B_0 = 1.19703\,{\rm T}$ \cite{GSI2}.  The terms,
proportional to $(1 - 1/\gamma)$, come from the Thomas precession
\cite{JJ83}. The electric charges of the interacting particles are
defined in terms of the electric charge of the proton $e$.

For the calculation of the amplitude of the $EC$--decays $m \to d +
\nu_e$ of the mother H--like heavy ion $m$ we have to use a standard
weak interaction Hamilton operator
\begin{eqnarray}\label{label7}
 \hspace{-0.15in}&&H_W(t) = \frac{G_F}{\sqrt{2}} V_{ud} \int
 d^3x [\bar{\psi}_n(x)\gamma^{\mu}(1 - g_A\gamma^5)
 \psi_p(x)]\nonumber\\
\hspace{-0.15in}&&\times\,[\bar{\psi}_{\nu_e}(x) \gamma_{\mu}(1 -
 \gamma^5)\psi_e(x)]
\end{eqnarray}
with standard notations \cite{Ivanov1}. As has been shown in
\cite{Ivanov2}, the non--trivial contribution to the $EC$--decay rate
of the H--like heavy ion in the ground $(1s)_{F = \frac{1}{2}, M_F \pm
\frac{1}{2}}$ comes from the state with the wave function
$|t,(1s)_{\frac{1}{2},-\frac{1}{2}}\rangle$. This means that the
evolution of the H--like heavy ion into the state $d + \nu_e$ is
defined by the wave function
$|t,(1s)_{\frac{1}{2},-\frac{1}{2}}\rangle$ only.
 
In the laboratory frame the evolution of the mother H--like ion $m$ in
time is described by the wave function $|t, (1s)_{\frac{1}{2},-\frac{1}{2}}\rangle$
\begin{eqnarray}\label{label8}
\hspace{-0.15in}|t, (1s)_{\frac{1}{2},-\frac{1}{2}}\rangle &=& -
e^{\,-iE^{(-+)}_mt} \sqrt{\frac{2}{3}}\,|1,- 1\rangle|\frac{1}{2}, +
\frac{1}{2}\rangle\nonumber\\
\hspace{-0.15in}&& +
e^{\,-iE^{(0-)}_m t}\sqrt{\frac{1}{3}}\,|1,0\rangle|\frac{1}{2}, -
\frac{1}{2}\rangle,
\end{eqnarray}
where $|I,I_z\rangle$ and $|s,s_z\rangle$ are spinorial wave functions
of the nucleus and the electron of the H--like heavy ion with
eigenvalues $I = 1, I_z = 0,\pm1$ and $s = \frac{1}{2}, s_z = \pm
\frac{1}{2}$, respectively. The energies $E^{(-+)}_m$ and $E^{(0-)}_m$
are defined by
\begin{eqnarray}\label{label9}
&&E^{(-+)}_m = E_m + \Big(a_e +
\frac{1}{\gamma}\Big)\,\mu_B\,B_0\,\cos\theta_e\nonumber\\ &&+
\Big(g_I - \frac{2 Z m_p}{M_I}\Big(1 -
\frac{1}{\gamma}\Big)\Big)\,\mu_N\,B_0\,\cos\theta_I,\nonumber\\ &&
E^{(0-)}_m = E_m - \Big(a_e +
\frac{1}{\gamma}\Big)\,\mu_B\,B_0\,\cos\theta_e,
\end{eqnarray}
where $E_m = \gamma M_m - i\,\frac{1}{2}\,\lambda_m$ and $\lambda_m$
is the weak decay rate of the H--like mother ion in the laboratory
frame \cite{PDG08}, $\theta_e$ and $\theta_I$ are the angles between
the z--axis and the axes of quantisation of the spins of the electron
and the nucleus, respectively. Since in the GSI experiments the
H--like heavy ions are in the $(1s)_{F = \frac{1}{2}}$, the spins of
the electron and the nucleus should be anti--parallel. This implies
that $\cos\theta_e = - \cos\theta_I$.

Using the wave function $|t,(1s)_{\frac{1}{2},- \frac{1}{2}}\rangle$
the probability of finding a mother H--like heavy ion at time $t$ is
\begin{eqnarray}\label{label10}
\hspace{-0.3in}&&P_m(t;\theta_e) = e^{-\lambda_mt}\,|\langle
(1s)_{\frac{1}{2},-\frac{1}{2}},
t|0,(1s)_{\frac{1}{2},-\frac{1}{2}}\rangle|^2 = \nonumber\\
\hspace{-0.3in}&&= e^{-\lambda_mt}\,\frac{5}{9}\,\Big(1 +
\frac{4}{5}\cos(\omega_B \cos\theta_e t)\Big),
\end{eqnarray}
where the frequency $\omega_B$ is equal to
\begin{eqnarray}\label{label11}
\hspace{-0.3in}&&\omega_B = \Big[\Big(2 a_e + \frac{2}{\gamma}\Big)
  - \Big(g_I - \frac{2 Z m_p}{M_I}\Big(1 -
  \frac{1}{\gamma}\Big)\Big)\frac{m_e}{m_p}\Big]\nonumber\\
\hspace{-0.3in}&&\times \mu_B B_0 = 1.34\times 10^{11}\,{\rm s^{-1}}.
\end{eqnarray}
Due to the factor $m_e/m_p$ the dominant contribution comes from the
electron anomalous magnetic moment. A period of the time modulation is
equal to
\begin{eqnarray}\label{label12}
T_B = \frac{2\pi}{\omega_B} = 4.70\times 10^{-11}\,{\rm
s}.
\end{eqnarray}
Practically, the period $T_B$ is proportional to the electron mass.
This disagrees with the experimental $A$--scaling of the period of the
time modulation of the H--like ions, measured at GSI
\cite{GSI4}--\cite{GSI6}.  Such a period of the time modulation cannot
be measured at the present level of the experimental time resolution
at GSI \cite{Ivanov5}.
\begin{figure}[t]
\includegraphics[width = 0.80\linewidth]{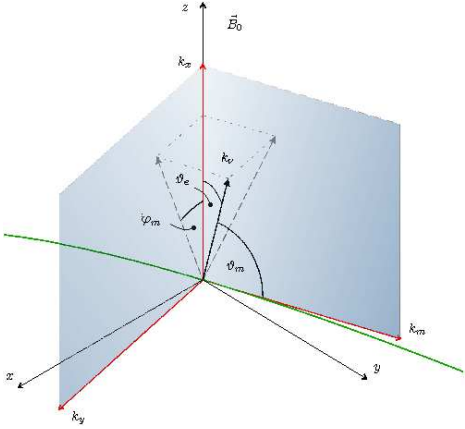}
\caption{Kinematical relations between the angles of the axis of
quantisation of spins and momenta of interacting particles.}
\end{figure}
The probability $P_m(t;\theta_e)$ should be averaged over the neutrino
angular distribution, calculated in the laboratory frame and given by
\begin{eqnarray}\label{label13}
\frac{dW_{\nu_e}}{d\Omega_m} = \frac{1}{8\pi \gamma^4}\,\frac{1}{(1 -
  v_m\cos\theta_m)^3},
\end{eqnarray}
where $d\Omega_m = \sin\theta_md\theta_md\varphi_m$ is an element of
the solid angle in the momentum space with axial axis directed along
the momentum $\vec{k}_m$ of the mother ion.  This gives
\begin{eqnarray}\label{label14}
\hspace{-0.3in}&&P_m(t) = \int
P_m(t;\theta_e)\,\frac{dW_{\nu_e}}{d\Omega_m}\,d\Omega_m =\nonumber\\
\hspace{-0.3in}&&= e^{-\lambda_mt}\,\frac{5}{9}\,\frac{1}{8\pi
\gamma^4}\int^{2\pi}_0\int^{\pi}_0
\frac{\sin\theta_md\theta_md\varphi_m}{(1 - v_m\cos\theta_m)^3}\nonumber\\
\hspace{-0.3in}&&\times\,\Big(1 + \frac{4}{5} \cos(\omega_B
\sin\theta_m\cos\varphi_m t)\Big).
\end{eqnarray}
As it is shown in Fig.\,1, the angle $\theta_e$ is related to angles
$\theta_m$ and $\varphi_m$ as follows $\cos\theta_e = \sin\theta_m
\cos\varphi_m$. Integrating over the azimuthal angle $\varphi_m$ and
using the integral representation Bessel functions and the properties
of infinite series of Bessel functions \cite{HMF72}  we get
\begin{eqnarray}\label{label15}
P_m(t) &=& e^{-\lambda_mt}\Big(1 -
\frac{4}{9}\frac{1}{\gamma^4}\int^{\pi}_0\frac{d\theta_m \sin\theta_m}{(1 -
v_m\cos\theta_m)^3}\nonumber\\
&&\times\,\sum^{\infty}_{n =
1}J_{2n}(\omega_B\sin\theta_m t)\Big),
\end{eqnarray}
This shows that the interaction of the mother H--like heavy ion with
the uniform magnetic field of the storage ring cannot provide a
time--modulation of the $EC$--decay rate, observed at the experiment
\cite{GSI2}.

We are grateful to T. Ericson for fruitful discussions.

\end{document}